\documentclass[journal]{IEEEtran}
\usepackage{amsmath,amsfonts}
\usepackage{algorithmic}
\usepackage{array}
\usepackage{textcomp}
\usepackage{stfloats}
\usepackage{url}
\usepackage{verbatim}
\usepackage{graphicx}
\def\BibTeX{{\rm B\kern-.05em{\sc i\kern-.025em b}\kern-.08em
    T\kern-.1667em\lower.7ex\hbox{E}\kern-.125emX}}
\usepackage{balance}

\usepackage{subfigure}
\usepackage{booktabs}
\usepackage{bm}
\usepackage{multirow}
\usepackage[linesnumbered,ruled]{algorithm2e}
\usepackage{orcidlink}
\usepackage{xcolor} 

\definecolor{ccr}{RGB}{0,0,255}  
\usepackage{hyperref}
\hypersetup{hypertex=true,
	colorlinks=true,
	linkcolor=ccr,
	urlcolor=black,
	citecolor=ccr,
	urlcolor=magenta}

\definecolor{mylightblue}{RGB}{100,120,250}

\begin{document}
\title{Exploring Distortion Prior with Latent Diffusion Models for Remote Sensing Image Compression}
\author{Junhui Li${^{\orcidlink{0000-0001-9143-1321}}}$, Jutao Li, Xingsong Hou${^{\orcidlink{0000-0002-6082-0815}}}$, and Huake Wang${^{\orcidlink{0000-0002-6548-5462}}}$
	\thanks{Manuscript received xx, 2024; This work was supported by the National Natural Science Foundation of China under Grant 62272376.  \textit{(Corresponding authors: Xingsong Hou.)}
		
	Junhui Li, Jutao Li, Xingsong Hou, and Huake Wang are with the School of Information and Communications Engineering, Xi’an Jiaotong University, Xi’an 710049, China (e-mail: mlkkljh@stu.xjtu.edu.cn, houxs@mail.xjtu.edu.cn).	
}}

\markboth{Journal of \LaTeX\ Class Files,~Vol.~18, No.~9, September~2020}%
{How to Use the IEEEtran \LaTeX \ Templates}

\maketitle

\begin{abstract}
	Learning-based image compression algorithms typically focus on designing encoding and decoding networks and improving the accuracy of entropy model estimation to enhance the rate-distortion (RD) performance. However, few algorithms leverage the compression distortion prior from existing compression algorithms to improve RD performance. In this paper, we propose a latent diffusion model-based remote sensing image compression (LDM-RSIC) method, which aims to enhance the final decoding quality of RS images by utilizing the generated distortion prior from a LDM. Our approach consists of two stages. In Stage I, a self-encoder learns prior from the high-quality input image. In Stage II, the prior is generated through a LDM, conditioned on the decoded image of an existing learning-based image compression algorithm, to be used as auxiliary information for generating the texture-rich enhanced images. To better utilize the prior, a channel attention and gate-based dynamic feature attention module (DFAM) is embedded into a Transformer-based multi-scale enhancement network (MEN) for image enhancement. Extensive experimental results demonstrate the proposed LDM-RSIC outperforms existing state-of-the-art traditional and learning-based image compression algorithms in terms of both subjective perception and objective metrics.  The code will be available at \textit{\url{https://github.com/mlkk518/LDM-RSIC}}.
\end{abstract}


\section{Introduction}
\IEEEPARstart{W}{ith} the ongoing development of remote sensing (RS) technology,  the volume of RS images is growing dramatically \cite{han2023edge}. This mainly stems from the continuous upgrading of platforms such as satellites and airplanes, as well as the wide application of high-resolution sensors \cite{pan2023coupled, 10516594}. Compression can help reduce the cost of data storage and transmission, and improve the efficiency of data processing and analysis, especially important in tasks that use high-resolution RS images for real-time monitoring. Moreover, with the wide application of RS images in various applications, there is an increasing demand for fast data acquisition, sharing, and processing. Consequently, the necessity to compress RS images is becoming increasingly imperative. 

Conventional image compression methods like JPEG2000 \cite{taubman2002jpeg2000}, BPG \cite{bpg2017}, and VVC \cite{VVC2021} have been pivotal in facilitating the storage and transmission of image data. However, these established standards suffer from two notable drawbacks \cite{pan2023hybrid, cheng2020learned}. Firstly, the encoding or decoding processes necessitate a sequential block-by-block implementation in block-based hybrid codes, leading to undesired blocking or ringing artifacts in the decoded images. Secondly, the intricate interdependencies among hand-crafted modules make it challenging to jointly optimize the entire coding algorithm. Furthermore, the growing demand for high-resolution RS images and the emergence of diverse applications with specific requirements have led to the exploration of advanced compression methods.


With the development of deep learning technology, learning-based image compression algorithms have made great progress \cite{10372532, 47602, he2022elic, 2021ITIP_liu, 10091784, fu2024weconvene}. Notably, Ballé \textit{et al.} \cite{47602} first introduced additional side information in the form of a hyperprior entropy model to estimate a zero-mean Gaussian distribution, highly improving the rate-distortion (RD) performance.  Cheng \textit{et al.} \cite{cheng2020learned} further enhanced RD performance by utilizing discretized Gaussian mixture likelihoods to parameterize entropy model distributions. 
To obtain higher RD performance and running speed, He \textit{et al.} \cite{he2022elic} proposed an uneven channel-conditional adaptive grouping method to improve the prediction accuracy of the entropy model, and developed the efficient learned image compression (ELIC) algorithm. To further boost RD performance, Fu \textit{et al.} \cite{fu2024weconvene} employed discrete wavelet transform (DWT) in their compression network design to reduce frequency-domain correlations. However, although the above learning-based compression algorithms have achieved impressive RD performance in natural scenes, RS image compression has not made significant progress due to the rich texture, context, and spectral information in RS images \cite{xiang2023remote, lu2017exploring}. \par

To overcome this obstacle, Zhang \textit{et al.} \cite{zhang2023global} strengthened the network's feature extraction capabilities by introducing a multi-scale attention module. To improve the entropy model, they also added global priors and anchored-stripe attention. Pan \textit{et al.} \cite{pan2023coupled} employed generative adversarial networks (GANs) to separately decode image content and complex textures for effective low-bitrate RS image compression. Wang \textit{et al.} \cite{10251976} observed that the conventional method of acquiring satellite images only uses the downlink to send compressed data to ground stations. They proposed using the uplink to utilize historical ground station images as references for on-orbit compression, reducing redundancy in RS images and improving compression efficiency. Additionally, Xiang \textit{et al.} \cite{10379598} used DWT to separate image characteristics into high- and low-frequency components, then developed compression networks to accurately represent both types of characteristics for high compression efficiency. 

While the aforementioned algorithms have demonstrated remarkable RD performance in RS image compression, deriving text-rich decoded images, especially at low bitrates, remains challenging. Recently, latent diffusion models (LDMs) have garnered significant attention due to their powerful ability to balance complexity reduction and intricate detail preservation \cite{10296015, wu2023seesr, chen2024hierarchical, Corneanu_2024_WACV, 10462910}. Unlike traditional DMs \cite{wang2024sam, lugmayr2022repaint, NEURIPS2023CCf6d8b4}, which typically operate directly in pixel space and require extensive computational resources for model optimization, LDMs can train models with limited computational resources while retaining high quality and flexibility \cite{rombach2022high}. For example, Chen \textit{et al.} \cite{chen2024hierarchical} deployed a DM in latent space (\textit{i.e.}, LDM) to generate prior, which were then embedded into an enhanced network through a hierarchical integration module for image deblurring. Similarly, Corneanu \textit{et al.} \cite{Corneanu_2024_WACV} performed forward and backward fusion steps using LDM, achieving impressive performance in terms of both image inpainting and running efficiency.

In this paper, we propose a novel LDM-based remote sensing image compression (LDM-RSIC) method.  Specifically, our approach comprises two main stages:
In Stage I, we develop a self-encoder to learn the compression distortion prior information from an existing image compression algorithm, ELIC. Prior information here refers to a compact representation of the residual or distortion between the original and compressed images, which helps the model understand and correct compression-induced detail loss. This prior encapsulates knowledge about the specific types of distortion commonly introduced during compression, particularly at low bitrates, such as loss of texture details or sharpness. By learning this prior, the model gains the ability to predict and mitigate these distortions during the decoding process.
 In Stage II, we generate this prior information using a LDM, conditioned on the decoded images. This generated prior information serves as auxiliary data and is fed into a multi-scale enhancement network (MEN) to enhance the quality of the decoded images. The key idea is that by utilizing the prior information, the MEN can better recover fine textures and structural details that may have been lost during the initial compression. 
 Furthermore, a channel and gate-based dynamic feature attention module (DFAM) is embedded into the MEN for better integration of the prior information with the decoded images.

The primary contributions of this paper can be summarized as follows:
\begin{itemize}
	\item[$\bullet$] We propose LDM-RSIC for RS image compression, leveraging the power of LDM to generate compression distortion prior, which is then utilized to enhance the image quality of the decoded images. The proposed LDM-based scheme can be adopted to improve the RD performance of both the learning-based and traditional image compression algorithms.
	\par
	\item[$\bullet$] We employ LDM and develop forward-backward fusion steps on the latent space to generate prior information instead of the pixel space for RS image compression. This approach not only saves on expensive training but also reduces inference time.
	\par
	\item[$\bullet$] A channel attention and gate-based DFAM is embedded in a Transformer-based MEN, facilitating dynamic fusion between the features of decoded images and the prior information. \par 
	\item[$\bullet$] Extensive experiments conducted on two RS image datasets demonstrate the superior performance of LDM-RSIC compared to state-of-the-art traditional and learning-based image compression algorithms. 
	\par
\end{itemize}

The remainder of the paper is structured as follows. Section \ref{Sec: realated} provides a review of related work on learning-based image compression and DMs. Section \ref{Sec:Method} elaborates on the proposed LDM-RSIC. Section \ref{Sec: Experiments} presents the experimental results and analysis conducted to evaluate the performance of the proposed method. Finally, Section \ref{Sec:Conclusion} concludes the paper.

\section{Related Work} \label{Sec: realated}
In this section, we present related work from two perspectives. Firstly, we review the realm of learning-based image compression, an emerging field that has garnered significant attention in recent years. Secondly, we explore the DMs, which serve as a key source of inspiration for the proposed LDM-RSIC.

\subsection{Learning-based Image Compression}
In recent years, the rapid advancement of deep learning has promoted the development of numerous learning-based image compression techniques \cite{qian2020learning, zou2022devil, he2022elic, pan2023coupled, zhang2023global}. These advancements build upon the pioneering work by Ballé \textit{et al.} \cite{47602}, who introduced a hyperprior entropy model to estimate zero-mean Gaussian distributions, facilitating compact latent representation within an end-to-end variational autoencoder-based framework \cite{balle2017end}.

When developing learning-based image compression methods, there are typically two key considerations. First, the redundancy coefficients among latent representations are reduced by improving the encoder network to save coding streams. Second, more accurate entropy models are designed to accurately estimate the probability distribution of these coefficients, thus better controlling the bitrate required for potential representation coding \cite{hu2021learning}.

In natural image compression, many efforts have been dedicated to achieving compact representation through the refinement of encoder and decoder networks \cite{zhang2023global, tang2022joint, xie2021enhanced, gao2021neural}. For instance, Gao \textit{et al.} \cite{gao2021neural} proposed a back projection scheme with attention and multi-scale feature fusion. Tang \textit{et al.} \cite{tang2022joint} employed asymmetric convolutional neural networks and attention mechanisms to enhance RD performance. Also, accurate entropy model development has garnered wide attention \cite{47602, qian2022entroformer, lee2018context, qian2020learning, cheng2020learned, he2022elic, 9067005}. Specifically, Cheng \textit{et al.} \cite{cheng2020learned} introduced discretized Gaussian mixture likelihoods for entropy model parameterization. Qian \textit{et al.} \cite{qian2022entroformer} proposed a Transformer-based entropy model by leveraging the Transformer's long-range dependency capabilities. Moreover, He \textit{et al.} \cite{he2022elic} developed an uneven channel-conditional adaptive grouping scheme to enhance coding performance without compromising speed. Liu \textit{et al.} \cite{liu2023learned} incorporated the local modeling ability of CNN and the non-local modeling ability of Transformers and developed a Transformer-CNN Mixture (TCM) block, which was adopted to improve the RD performance.  

In RS image compression, several studies focused on exploring competitive network structures and image transformations \cite{pan2023coupled, 10379598, pan2023hybrid, zhang2023global, NEURIPS2023CCf6d8b4}. For example, Pan \textit{et al.} \cite{pan2023coupled} utilized GANs to independently decode image content and detailed textures, thereby enhancing compression performance at low bitrates. Xiang \textit{et al.} \cite{10379598} leveraged DWT to enhance the representation of high and low-frequency features. Li \textit{et al.} \cite{10516594} introduced Transformer and patch-based local attention modules to develop a competitive encoder network and entropy model for object-fidelity RS image compression. Pan \textit{et al.} \cite{pan2023hybrid} proposed a hybrid attention network to improve entropy model prediction accuracy. Zhang \textit{et al.} \cite{zhang2023global} employed a multi-scale attention module and global priors to enhance feature extraction and improve the entropy model. 
More recently, Yang \textit{et al.} \cite{NEURIPS2023CCf6d8b4} encoded inputs image into compact representtaions, which were used as conditions for a DM. However, as the involved ``texture" are synthesized on the fly, this conditions significantly affect the final image quality of the decoded images, leading to poor RD performance.

In contrast to these approaches, our objective is to employ conditional LDM to generate compression distortion prior induced by one of the existing compression algorithms. Subsequently, this prior information is utilized to enhance the RD performance of the compression algorithm.

\subsection{Diffusion Models (DMs)}
DMs \cite{10081412, zhao2024uni, 10081412}, as probabilistic generative models, utilize parameterized Markov chain to optimize the lower variational bound on the likelihood function. This enables the construction of desired data samples from Gaussian noise via a stochastic iterative denoising process. Recently, DMs have become increasingly influential in image restoration tasks,  such as image super-resolution \cite{wang2024sam}, inpainting \cite{lugmayr2022repaint}, and deblurring \cite{ren2023multiscale}. 
However, since these models typically operate directly in pixel space, optimizing the most powerful DMs often requires hundreds of GPU days, and the cost of inference is also high due to sequential evaluation \cite{rombach2022high}. 

To train DMs on limited computational resources while maintaining their quality and flexibility, Rombach \textit{et al.} \cite{rombach2022high} innovatively applied them to the latent space of powerful pretrained autoencoders, thus developing the LDM. Compared to previous DM-based methods, LDM achieves a near-optimal balance between reducing complexity and preserving details for the first time, greatly enhancing visual fidelity. Subsequently, a series of LDM-based studies have been reported in a wide range of applications \cite{wu2023seesr, chen2024hierarchical, Corneanu_2024_WACV}. For example, in \cite{chen2024hierarchical}, the authors leveraged LDM to generate prior features, which were then fused to an enhancement network by a hierarchical integration module for image deblurring. Besides, in \cite{Corneanu_2024_WACV}, the authors resorted to the powerful ability of LDM and developed an impressive LDM-based image inpainting algorithm.

\section{Methodology} \label{Sec:Method}
In this section, the proposed LDM-RSIC and its two novel designs will be introduced in detail. We will first begin with an overview of LDM-RSIC. Subsequently, two novel designs, namely, prior information learning and LDM-based prior information generation, will be presented.

\subsection{Overview of LDM-RSIC}
\begin{figure*}[!htbp]
	\centering
	\includegraphics[width=0.98 \textwidth]{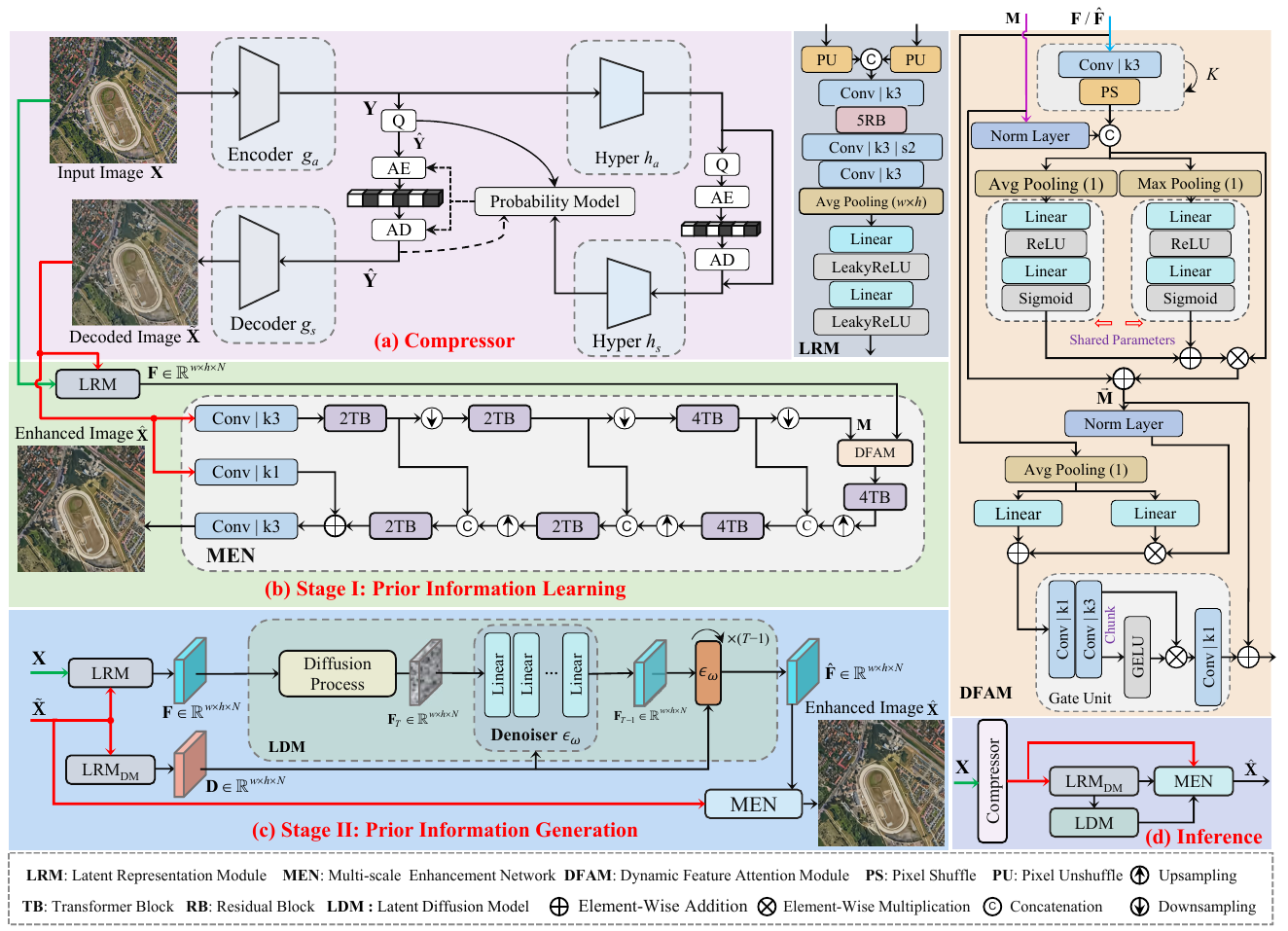}
	\caption{Overview of the proposed LDM-RSIC, which comprises the compressor, MEN, and LDM. The compressor utilizes the competitive image compression algorithm ELIC \cite{he2022elic}. ``2TB" indicates two stacked Transformer blocks, and ``5RB" denotes five serially connected residual blocks. ``AE" and ``AD" refer to the arithmetic encoder and decoder, respectively. Stage I aims to learn the prior information $\mathbf{F}$, and Stage II focuses on employing LDM to generate the prior features $\mathbf{\hat{F}}$ to replace $\mathbf{F}$.}
	\label{Fig:main_framework}
\end{figure*}

The main framework of the proposed LDM-RSIC is depicted in Fig. \ref{Fig:main_framework}, where ``Conv $ | $ k3 $ | $ s2 " refers to the convolution layer with a kernel size of $3 \times 3$ and stride 2. By default, the stride value is set to 1. This framework primarily comprises three components: the compressor, multi-scale enhancement network (MEN), and latent diffusion module (LDM). \par
Specifically, the compressor refers to an existing image compression algorithm, here the competitive algorithm ELIC is employed as the compressor. Given an input image $\mathbf{X} \in \mathbb{R}^{ W \times H \times 3}$, the encoder $ g_a $ aims to extract the latent features $ \mathbf{Y} $ in a compact manner. These features $ \mathbf{Y} $ are then quantized by a quantification Q to attain $\mathbf{\hat Y}$. By encoding $\mathbf{\hat Y}$ with a hyper entropy coder, the image can be compressed into a data stream. On the decompression side, the reconstructed image can be obtained using  $ \mathbf{\hat Y }$ and the decoder network $ g_s$. The above process can be formulated as
\begin{equation}
	\begin{aligned}
		\mathbf{Y} &= g_a ( \mathbf{X}; \phi ), \\
		\mathbf{\hat Y} &= \operatorname{Q} ( \mathbf{Y} ), \\
		{{\Tilde{\mathbf{X}}}} &= g_s ( \mathbf{\hat Y}; \varphi ),
	\end{aligned}
\end{equation}
where the encoder $ g_a $ and decoder $ g_s $ represent neural networks developed in \cite{he2022elic}, with $\phi$ and $\varphi$ denoting their network parameters, respectively. $\operatorname{Q}(\cdot)$ denotes a quantization operation. To encode the latent $\mathbf{\hat Y}$, hyper networks $h_a$ and $h_s$ are typically developed to obtain the probability model, which is commonly used for arithmetic coding. For achieving different compression rates, the compression networks are often optimized using the loss function defined as
\begin{align}
	\mathcal{L}_\text{com} = \operatorname{R} + \lambda \operatorname{D} ( \mathbf{X} , g_s ( \mathbf{\hat { Y } ) }),
\end{align}
where the rate $\operatorname{R}$ represents the entropy calculation function of the quantized latent representation of the compressor, and $\operatorname{D}(\cdot)$ denotes the mean squared error (MSE). $\lambda$ is a hyperparameter used to control the balance between the rate and the quality of the decoded images, where a higher value of $\lambda$ indicates better quality of the decoded image and higher bitrates. \par
Over the past few years, numerous studies have focused on designing the network structures of $g_a$, $g_s$, $h_a$, and $h_s$. However, it seems challenging to further improve RD performance solely through network design. Considering the impressive ability of LDMs to generate rich information conditioned on known knowledge, we aim to employ the LDM to generate prior information, which is then embedded into MEN to enhance the quality of the decoded images, thus improving the RD performance of the compressor. This process can be achieved through two stages: prior information learning and prior information generation. Next, we will illustrate the two stages in detail.


\subsection{Stage I: Prior Information Learning}
In Stage I, as illustrated in Fig. \ref{Fig:main_framework}(b), both the input image $\mathbf{X}$ and the decoded image $\tilde{\mathbf{X}}$ are fused through the latent representation module (LRM) to obtain latent representation features $\mathbf{F}$. The features $\mathbf{F}$, enriched with high-quality prior information, are then utilized to assist the MEN in reconstructing the enhanced image $\mathbf{\hat X}$. \par
\subsubsection{Latent Representation Module (LRM)}
To be specific, as depicted in the LRM component of Fig. \ref{Fig:main_framework}, the input image and decoded image undergo a pixel unshuffle (PU) operation to downsample the features. Subsequently, these downsampled features are concatenated and fed into several convolutional and residual block (RB) layers, followed by an average pooling layer with an output size of $w\times h$ to reduce the dimensionality. Afterward, two linear layers with LeakyReLU activation functions are employed to mixture the features. This process yields the latent representation given as
\begin{equation} \label{Eq:prior_learning}
	\mathbf{F} = \operatorname{LRM}({{\Tilde{\mathbf{X}}}}, \mathbf{X}),
\end{equation}
where the resolution of $\mathbf{F}$ is $ w\times h \times N$. Importantly, the token number $w\times h \times N$ remains a constant much smaller than $ W \times H \times 3 $. Consequently, the computational burden of the subsequent LDM is effectively reduced. In addition, it should be noted that each ``Conv" in LRM of Fig. \ref{Fig:main_framework} is followed by a LeakyReLU activation function. \par

\subsubsection{Multi-Scale Enhance Network (MEN)}
Considering the impressive performance of multi-scale Transformer-based networks in various tasks \cite{9785614, 10377629, chen2024hierarchical}, MEN is developed based on Transformer blocks (TBs) for the quality improvement of the decoded images. Furthermore, to better utilize the latent features $\mathbf{F}$, a channel attention and gate-based dynamic feature attention module (DFAM) is embedded in MEN.

In the development of DFAM, we first upsample the features $\mathbf{F}$ through $K$ stacked layers comprising a convolutional layer followed by a pixel shuffle (PS) layer. Subsequently, the MEN's deep features $\mathbf{M}$, after a layer normalization operation, are concatenated with the upsampled features along the channel dimension. As different channels hold varying importance for the final reconstruction, channel attention is introduced to dynamically weigh the concatenated channels. Thus, as depicted in Fig. \ref{Fig:main_framework}, the output features $\Vec{\mathbf{M}}$ can be expressed as
\begin{equation}
	\begin{aligned}
		\mathbf{Z} &= \operatorname{Concat}\left(\operatorname{UP}\left(\mathbf{F}\right), \operatorname{Norm}\left(\mathbf{M}\right)\right),\\
		\Vec{\mathbf{M}} &= (\operatorname{FC}(\operatorname{AP}(\mathbf{Z})) + \operatorname{FC}(\operatorname{MP}(\mathbf{Z}))) \otimes \mathbf{Z} + \mathbf{M},
	\end{aligned}
\end{equation}
where $\operatorname{UP}(\cdot)$ denotes the upsampling operation function described above; $\operatorname{FC}(\cdot)$, $\operatorname{AP}(\cdot)$, and $\operatorname{MP}(\cdot)$ represent the full connection layer, average pooling, and max pooling, respectively. $\otimes$ indicates element-wise multiplication.\par
Additionally, since not all learned latent features $\mathbf{F}$ contribute equally to the improvement of the reconstruction, we incorporate a gate adjuster to further reweight the obtained features $\Vec{\mathbf{M}}$. Specifically, before further processing, the prior information $\mathbf{F}$ adjusts the features $\Vec{\mathbf{M}}$ through scaling and shifting operations, followed by a gate unit (GU) to selectively amplify or suppress certain features based on their relevance. Therefore, the output of DFAM can be derived as
\begin{equation}
	\Vec {\Vec{\mathbf{M}}} =   \operatorname{GU}(\operatorname{Norm}(\vec{\mathbf{M}}) \otimes \operatorname{LL}(\operatorname{AP}(\mathbf{F})) + \operatorname{LL}(\operatorname{AP}(\mathbf{F})))+ \Vec{\mathbf{M}},
\end{equation}
where $\operatorname{LL}(\cdot)$ refers to a linear layer. This integration allows the model to adjust $\Vec{\mathbf{M}}$ based on the prior information $\mathbf{F}$, which enhances the flexibility and adaptability of the model.
\subsubsection{Prior Learning Loss}
To obtain the latent features $\mathbf{F}$, the pretrained model parameters of ELIC are frozen, and MEN and LRM are jointly optimized using the L1 norm given as
\begin{align}
	\mathcal{L}_\text{learn}({\mathbf{\Omega}_1}) = \Vert \operatorname{MEN}({{\Tilde{\mathbf{X}}}}, \operatorname{LRM}({{\Tilde{\mathbf{X}}}}, \mathbf{X})) - \mathbf{X} \Vert _ { 1 },
\end{align}
where ${{\Tilde{\mathbf{X}}}}$ is the decoded image of ELIC with a specific value of $\lambda$. ${\mathbf{\Omega}_1}$ refers to the trained network parameters. Thus, the process of stage I can be summarized in Algorithm \ref{alg:A}.

\begin{algorithm} [!htbp] 
	\caption{Process of stage I}  	
	\label{alg:A}  	
	\hspace*{0.2in}\small{Input:}
	$ \mathbf{X}, \lambda $ \\
	\hspace*{0.2in}\small{Output:} 
	${\mathbf{\Omega}_1}, \mathbf{F}$
	\begin{algorithmic}[1] 
		\STATE  {\small {\textbf{Prior Information Learning:}  }}
			\begin{ALC@g}
				\STATE  \small{Freeze the pretrained ELIC.}
				\FOR{$\text{epoch} \gets 1$ \textbf{to} $\text{epochs}1$ } 
				\STATE \small{${{\Tilde{\mathbf{X}}}} = \operatorname{ELIC}(\mathbf{X}, \lambda), $}
				\STATE	\small{\textcolor{mylightblue}{/\//\ Learned prior features:}} \\
				\STATE \small{$	\mathbf{F} = \operatorname{LRM}({{\Tilde{\mathbf{X}}}}, \mathbf{X}),$}				
				\STATE	\small{\textcolor{mylightblue}{/\//\ Compute prior learning loss:}} \\
				$\mathcal{L}_\text{learn}({\mathbf{\Omega}_1}) = \Vert \operatorname{MEN}({{\Tilde{\mathbf{X}}}}, \mathbf{F}) - \mathbf{X} \Vert _ { 1 }$,
				\STATE \small{\textcolor{mylightblue}{/\//\  Update ${\mathbf{\Omega}_1}$ using gradient descent:}} \\
				$\small {{\mathbf{\Omega}_1} \gets {\mathbf{\Omega}_1} - \alpha_1 \nabla \mathcal{L}_{\text{learn}}({\mathbf{\Omega}_1})},$	
				\ENDFOR	
			\end{ALC@g}
			\STATE {\small {\textbf{Return}}: ${\mathbf{\Omega}_1}$}	 
		\end{algorithmic}
	\end{algorithm}
	
\subsection{Stage II: Prior Information Generation}
In Stage II, the LDM is trained to generate the prior features $\mathbf{\hat F}$, which replace the features $\mathbf{F}$ of MEN for higher enhanced image quality. Concretely, the LDM is developed based on conditional DDPM \cite{NEURIPS2023_3ec077b4, 10479050, 10.1145/3589334.3645514}. The LDM involves a forward diffusion process and a reverse denoising process, as depicted in Fig. \ref{Fig:main_framework}(c).

\subsubsection{Diffusion Process}
Given an input image $\mathbf{X}$, we first employ the LRM trained in Stage I to generate the corresponding prior features $\mathbf{F} \in \mathbb{R}^{w\times h \times N}$. We take $\mathbf{F}$ as the starting point of the forward Markov process and gradually add Gaussian noise to it over $T$ iterations as follows:
\begin{equation}
	\begin{aligned}
		q ( \mathbf{F} _ { 1 : T} | \mathbf{F} _{0}) &= \prod _ { t = 1 } ^ { T } q ( \mathbf{F} _ { t } | \mathbf{F} _ { t - 1 } ) ,  \\
		q ( \mathbf{F} _ { t } | \mathbf{F} _ { t - 1 } ) &= \mathcal{N} ( \mathbf{F} _ { t } ; \sqrt { 1 - \eta _ { t } } \mathbf{F} _ { t - 1 } , \eta _ { t } \mathbf{I} ) ,
	\end{aligned} 
	\label{Eq:Add_noise}
\end{equation}
where $\mathbf{F}_t$ represents the noisy features at the $t$-th step and $\mathbf{F}_0 = \mathbf{F}$; $\eta_{1:T}\in(0, 1)$ are hyperparameters that control the variance of the noise; $\mathcal{N}$ denotes the Gaussian distribution. Based on the iterative derivation presented in \cite{kingmaauto}, we can reformulate Eq. (\ref{Eq:Add_noise}) as
\begin{align}
	q ( \mathbf{F} _ { T } | \mathbf{F} _ { 0 } ) = \mathcal{N} ( \mathbf{F} _ { T } ; \sqrt { \overline { \gamma }_{T} } \mathbf{F} _ { 0 } , ( 1 - \overline { \gamma } _ { T } ) \mathbf{I} ) ,  \label{Eq:Add_t_noise}
\end{align}
where  $\overline { \gamma } _ { t } = \prod _ { t = 1 } ^ { T }\gamma _ { t }$ and $\gamma _t = 1 - \eta _ { t }$.

\subsubsection{Reverse Process}
Here, our objective is to generate the prior features from a pure Gaussian distribution. The reverse process is a $T$-step Markov chain that runs backward from $\mathbf{F}_T$ to $\mathbf{F}_0$. Specifically, for the reverse step from $\mathbf{F}_t$ to $\mathbf{F}_{t-1}$, we use the posterior distribution given as
\begin{equation} \small
	\begin{aligned}
		p(\mathbf{F}_{t-1} | \mathbf{F}_t, \mathbf{F}_0) &= \mathcal{N}\left(\mathbf{F}_{t-1}; \mu_t(\mathbf{F}_t, \mathbf{F}_0), \frac{1 - \overline{\gamma}_{t-1}}{1 - \overline{\gamma}_t} \eta_t \mathbf{I}\right), \\ {s.t.}, ~~
		\mu_t(\mathbf{F}_t, \mathbf{F}_0) &= \frac{1}{\sqrt{\gamma_t}} \left(\mathbf{F}_t - \frac{1 - \gamma_t}{\sqrt{1 - \overline{\gamma}_t}} \epsilon\right),
	\end{aligned}
	\label{Eq:Reverse_processing}
\end{equation}
where $\epsilon$ represents the noise in $\mathbf{F}_t$, and is the only uncertain variable. Following previous works \cite{10377629, chen2024hierarchical}, we adopt a neural network, termed as denoiser ($\epsilon_\omega$), to estimate the noise $\epsilon$ for each step. Since the LDM operates in the latent space, we utilize another latent encoder, denoted as LRM$_\text{DM}$, which has the same structure as the network after LRM removes the single PU input branch. LRM$_\text{DM}$ compresses the decoded image ${{\Tilde{\mathbf{X}}}}$ into latent space to get the condition latent features $\mathbf{D} \in \mathbb{R}^{w\times h \times N}$. The denoising network predicts the noise conditioned on $\mathbf{F}_t$ and $\mathbf{D}$, \textit{i.e.}, $\epsilon_\omega(\mathbf{F}_t, \mathbf{D}, t)$. With the substitution of $\epsilon_\omega$ in Eq. (\ref{Eq:Reverse_processing}) and set the variance to $(1-\gamma_t)$, thus we can derive
\begin{align}
	\mathbf{F} _ { t - 1 } = \frac { 1 } { \sqrt { \gamma _ { t } } } ( \mathbf{F} _ { t } - \frac { 1 - \gamma _ { t } } { \sqrt { 1 - \overline { \gamma } _ { t } } } \epsilon _ { \omega } ( \mathbf{F} _ { t } , \mathbf{D} , t ) ) + \sqrt { 1 - \gamma _ { t } } \epsilon _ { t } ,   \label{Eq: Reverse_pro}
\end{align}
where ${\epsilon}_t \sim \mathcal{N}(0, \mathbf{I})$. By iteratively sampling $\mathbf{F}_t$ using Eq. (\ref{Eq: Reverse_pro}) $T$ times, we can generate the predicted prior features ${\mathbf{\hat{F}}} \in \mathbb{R}^{w\times h \times N}$, as depicted in Fig. \ref{Fig:main_framework}(c). The predicted prior features are then used to guide MEN. It should be noted that as the distribution of the latent features $\mathbf{F} \in \mathbb{R}^{w\times h \times N}$ is much simpler than that of the input image $ \mathbf{X} \in \mathbb{R}^{ W \times H \times 3} $, the prior features ${\mathbf{\hat{F}}}$ can be generated with a small number of iterations.

\subsubsection{Prior Generation Loss}
In this stage, the models of ELIC and LRM from Stage I are frozen, and the pretrained network parameters of LRM and MEN are respectively used to initialize the networks of LRM$_\text{DM}$ and MEN for Stage II. Training LDM involves training the denoising network $\epsilon _ \omega$. As adopted in the studies \cite{rombach2022high, ho2020denoising}, we train the model by optimizing the network parameters $\omega$ of $\epsilon_\omega$, which can be formulated as
\begin{align}
	\nabla _ { \omega } \Vert \epsilon - \epsilon _ { \omega } ( \sqrt { \overline{\gamma} } _ { t } \mathbf{F} + \sqrt { 1 - \overline { \gamma } _ { t } } \epsilon , \mathbf{D} , t ) \Vert _ { 2 } ^ { 2 } ,
\end{align}
where $\mathbf{F}$ and  $\mathbf{D}$ refer to the learned prior features and condition latent features derived above; $t \in [1, T]$ is a random time step, and $\epsilon \sim \mathcal{N}(0, \mathbf{I})$ denotes sampled Gaussian noise. 

For each training iteration, we use the prior features $\mathbf{F}$ to generate the noise sample $\mathbf{F}_T$ according to Eq. (\ref{Eq:Add_t_noise}). Given that the time-step $T$ is relatively small in latent space, we then run the complete $T$ iteration reverse processes (\textit{i.e.}, Eq. (\ref{Eq: Reverse_pro})) to generate the predicted prior features ${\mathbf{\hat{F}}}$, which can be derived by 
\begin{equation} \label{Eq: Generated_prior}
	\mathbf{\hat{F}} =  \operatorname{LDM}(\operatorname{LRM}_\text{DM}({{\Tilde{\mathbf{X}}}})).
\end{equation}
The obtained features  $\mathbf{\hat{F}}$ are then used to provide prior information for MEN. Thereafter, the loss function can be formulated as
\begin{equation} 
	\begin{split}
		\mathcal{L}_\text{gen}(\mathbf{\Omega_2}) = \Vert {\operatorname{MEN}}({\tilde{\mathbf{X}}},  \mathbf{\hat{F}}) - \mathbf{X} \Vert _{ 1 } 
		+    \Vert {\mathbf{\hat{F}}} - \mathbf{F}  \Vert _ {1}, 
	\end{split}
\end{equation}
where the first term on the right side of the equation is the quality fidelity item, and the second term is the diffusion loss.
\begin{algorithm} [!htbp] 
	\caption{Process of stage II}  	
	\label{alg:B}  	
	\hspace*{0.2in}\small{Input:}
	$ \mathbf{X}, {\mathbf{\Omega}_1}, \mathbf{F}, \lambda $ \\
	\hspace*{0.2in}\small{Output:} 
	$ {\mathbf{\Omega}_2}$ 	
	\begin{algorithmic}[1] 
				\STATE {\small {\textbf{Prior Information Generation:} }}
				\begin{ALC@g}
					\STATE  \small{Freeze the pretrained ELIC and LRM; \\ Initialize MEN and LRM$_\text{DM}$ with ${\mathbf{\Omega}_1}$.}
					\FOR{$\text{epoch} \gets 1$ \textbf{to} $\text{epochs}2$ } 
					\STATE	\small{\textcolor{mylightblue}{/\//\ Learned prior features:}} \\
					\STATE \small{$	\mathbf{F} = \operatorname{LRM}({{\Tilde{\mathbf{X}}}}, \mathbf{X}),$}	
					\STATE	\small{\textcolor{mylightblue}{/\//\ Generated prior features:}} \\
					\STATE \small{$	\mathbf{\hat{F}} =  \operatorname{LDM}(\operatorname{LRM}_\text{DM}({{\Tilde{\mathbf{X}}}})),$}
					\STATE	\small{\textcolor{mylightblue}{/\//\ Compute prior generation loss:}} \\
					\small{
	$	\mathcal{L}_\text{gen}({\mathbf{\Omega}_2})  = \Vert\operatorname{MEN}({{\Tilde{\mathbf{X}}}},  \mathbf{\hat{F}}) - \mathbf{X} \Vert_{ 1 }  	 +    \Vert \mathbf{\hat{F}} - \mathbf{F}  \Vert_ {1}, $
					} \\
					\STATE \small{\textcolor{mylightblue}{/\//\  Update ${\mathbf{\Omega}_2}$ using gradient descent:}} \\
					$\small {{\mathbf{\Omega}_2} \gets {\mathbf{\Omega}_2} - \alpha_2 \nabla \mathcal{L}_{\text{gen}}({\mathbf{\Omega}_2})},$ 	
					\ENDFOR	
				\end{ALC@g}
				\STATE {\small {\textbf{Return}}: ${\mathbf{\Omega}_2}$}	 
			\end{algorithmic}
			\vspace{-0.5\baselineskip} 
			\hrulefill   \\ 
			\hspace*{0.2in}\small{Input:}
			$ \mathbf{X}, {\mathbf{\Omega}_2}, \lambda $ \\
			\hspace*{0.2in}\small{Output:} 
			$  {\mathbf{\Hat{X}}}$ 
			\begin{algorithmic}[1] 
				\STATE  {\small {\textbf{Inference:}  }}
				\begin{ALC@g}
					\STATE  \small{Load pretrained ELIC.} \\
					\STATE \small{$ {{\Tilde{\mathbf{X}}}} = \operatorname{ELIC}(\mathbf{X}, \lambda),  $}
					\STATE	\small{{Load ${\mathbf{\Omega}_2}$ for  $\operatorname{LRM}_\text{DM} $, $ \operatorname{LDM} $, and $ \operatorname{MEN} $.}} \\
					\STATE \small{$	\hat{\mathbf{F}} =  \operatorname{LDM}(\operatorname{LRM}_\text{DM}({{\Tilde{\mathbf{X}}}}))$}
					\STATE {$	\hat{\mathbf{{X}}} = \operatorname{MEN}({{\Tilde{\mathbf{X}}}}, \hat{\mathbf{F}}; {\mathbf{\Omega}_2})$ }\\	
				\end{ALC@g}	 
				\STATE {\small {\textbf{Return}: ${\mathbf{\Hat{X}}}$}}
			\end{algorithmic} 
		\end{algorithm}
\subsubsection{Inference}
As depicted in Fig. \ref{Fig:main_framework}(d), after obtaining the trained model, the input image \(\mathbf{X}\) undergoes compression using the compressor, ELIC. Subsequently, the compressed image $\Tilde{\mathbf{X}}$ is fed into Eq. (\ref{Eq: Generated_prior}) to generate the prior features \({\mathbf{\hat{F}}}\). 
Following this, MEN utilizes \({\mathbf{\hat{F}}}\) to produce the enhanced image $\hat{\mathbf{{X}}}$ with rich texture details. This process can be expressed as
\begin{equation}
	\begin{aligned}
		{{\Tilde{\mathbf{X}}}} &= \operatorname{ELIC}(\mathbf{X}, \lambda),\\
		\hat{\mathbf{{X}}} &= \operatorname{MEN}({{\Tilde{\mathbf{X}}}}, {\mathbf{\hat{F}}}; {\mathbf{\Omega}_2}),
	\end{aligned}
\end{equation} 
where ${\mathbf{\hat{X}}}$ refers to the enhanced image. Thereafter, the process of stage II and inference can be summarized in Algorithm \ref{alg:B}.

\section{Experimental Results and Analysis} \label{Sec: Experiments}
\subsection{Experimental Settings}
\subsubsection{Datasets}
To evaluate the performance of the proposed LDM-RSIC, we conducted experiments using two RS image datasets: DOTA \cite{Ding_2019_CVPR} and UC-Merced (UC-M) \cite{yang2010bag}, as adopted in \cite{10516594, pan2023hybrid}. The DOTA dataset consists of 2,806 images, each with pixel resolutions ranging from $800\times800$ to $4000 \times 4000$, containing objects of various scales, orientations, and shapes. The UC-M dataset contains 21 classes of RS scenes and a total of 2,100 images, each with a resolution of $256 \times 256$. 

During training, we use the training dataset of DOTA for model training and employ several augmentation strategies, such as random horizontal flip, random vertical flip, and random crop. To evaluate the performance of the proposed method, we construct the testing set as follows: We randomly select 100 images from the testing dataset of DOTA, and each of these images is then cropped to a resolution of $256 \times 256$. Additionally, we select 20\% of the images from each category in the UC-M dataset.

\subsubsection{Implementation Details}
The experiments are conducted on an Intel Silver 4214R CPU with 2.40GHz and one NVIDIA GeForce RTX 3090 Ti GPU, using the PyTorch 1.13.1 framework with CUDA 11.7. ELIC \cite{he2022elic} is employed to compress the images of the training dataset. Specifically, five trained models with parameters $\lambda \in \{4, 8, 32, 100, 450\} \times 10^{-4}$ are used to separately generate the decoded images, which serve as input for training the developed model. Since different values of $\lambda$ correspond to varying levels of compression distortion in ELIC, we use the decoded image under a specific $\lambda$ to train a LDM-RSIC model. The values of $w$, $h$, $N$, and $K$ are set to 4, 4, 256, and 3, respectively.
Besides, we utilize the widely used Adam optimizer \cite{kingma2014adam} with $\beta_1 = 0.9$ and $\beta_2 = 0.999$ for model optimization. In Stage I, the initial learning rate $\alpha_1$ is set to $1 \times 10^{-4}$, and the total number of iterations is 300K. After 92K iterations, the model is trained with the initial learning rate gradually reduced to $1 \times 10^{-6}$ using cosine annealing \cite{loshchilov2017sgdr}. Early stopping is applied to prevent overfitting. In Stage II, the total epochs are set to 400K, with an initial learning rate of $\alpha_2 = 1 \times 10^{-4}$ for training the LRM$_\text{DM}$ and LDM before 200K iterations. Subsequently, the learning rate $\alpha_2$ is reduced to 0.1 times every 80K iterations to jointly finetune the MEN, LRM$_\text{DM}$, and LDM. The batch size is set to 4 during the two stages.

\subsubsection{Evaluation Metrics}
In this paper, we use bits per pixel (bpp) to evaluate coding bitrates, while peak signal-to-noise ratio (PSNR), multi-scale structural similarity (MS-SSIM), and learned perceptual image patch similarity (LPIPS) \cite{zhang2018unreasonable} serve as evaluation metrics. PSNR and MS-SSIM focus on numerical comparisons and structural similarity in images, while LPIPS emphasizes perceptual evaluation by leveraging deep learning-based models to extract high-level perceptual features. 

\begin{figure*}[!htbp]
	\centering
	\subfigure[\fontsize{8}{10}\selectfont Evaluation on DOTA]{
		\begin{minipage}[t]{0.33\textwidth}
			\centering
			\includegraphics[scale=0.78]{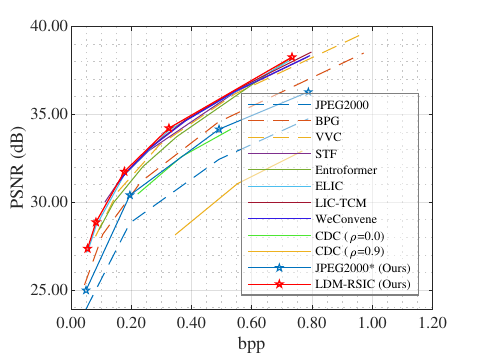}
		\end{minipage}%
		\begin{minipage}[t]{0.33\textwidth}
			\centering
			\includegraphics[scale=0.78]{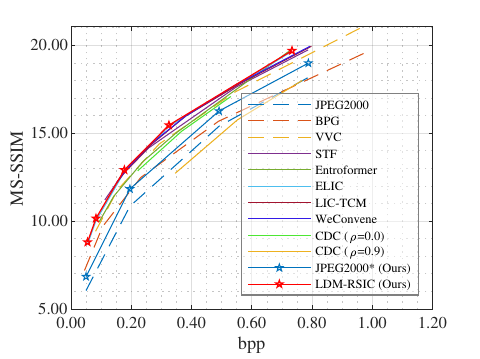}
		\end{minipage}
		\begin{minipage}[t]{0.33\textwidth}
			\centering
			\includegraphics[scale=0.78]{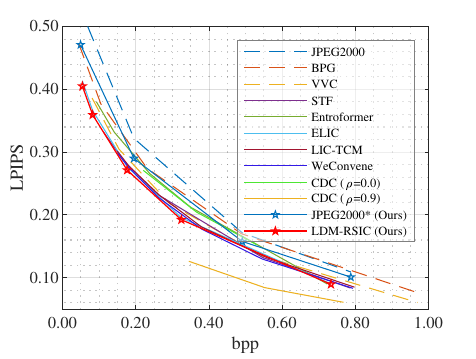}
		\end{minipage}
	}%
	\vspace{-3mm}
	\subfigure[\fontsize{8}{10}\selectfont Evaluation on UC-M]{
		\centering
		\begin{minipage}[t]{0.33\textwidth}
			\includegraphics[scale=0.78]{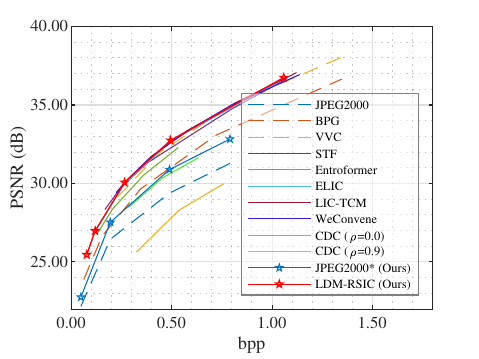}
		\end{minipage}
		\begin{minipage}[t]{0.33\textwidth}
			\includegraphics[scale=0.78]{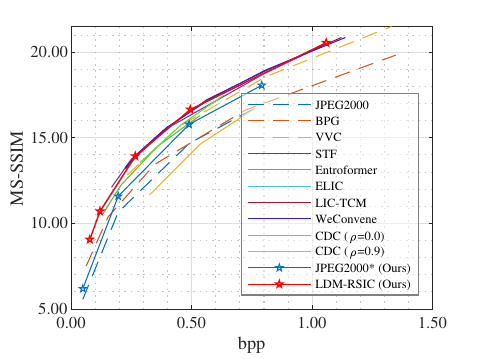}
		\end{minipage}
		\begin{minipage}[t]{0.33\textwidth}
			\includegraphics[scale=0.78]{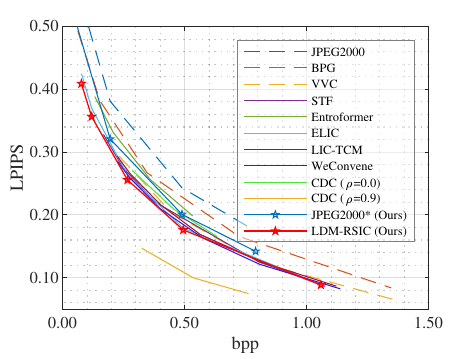}
		\end{minipage}
	}%
	\centering
	\caption{Performance evaluation on the testing sets of DOTA and UC-M in terms of PSNR, MS-SSIM, and LPIPS. LDM-RSIC and JPEG2000* refer to algorithms designed using ELIC and JPEG2000 as the compressor in Fig. \ref{Fig:main_framework}, respectively.}
	\label{Fig:RD_DOTA_UC}
\end{figure*}
\subsection{Quantitative Comparison}
To demonstrate the effectiveness of the proposed LDM-RSIC, we compare it with three traditional image compression standards, 
including JPEG2000 \cite{taubman2002jpeg2000}, BPG \cite{bpg2017} as well as VVC (YUV 444) \cite{VVC2021}, and the recent learning-based image compression algorithms, including Entrorformer \cite{qian2022entroformer}, STF \cite{zou2022devil},  ELIC \cite{he2022elic}, LIC-TCM \cite{liu2023learned}, 	CDC ($\rho\!=\!0.0$) \cite{NEURIPS2023CCf6d8b4}, CDC ($\rho\!=\!0.9$) \cite{NEURIPS2023CCf6d8b4}, and WeConvene \cite{fu2024weconvene}. In addition, we further replace the learning-based compressor (\textit{i.e.}, ELIC) in LDM-RSIC with the traditional algorithm JPEG2000, termed as JPEG2000*. Note that for CDC, the step number suggested by the authors was used in the experiments. \par

Figs. \ref{Fig:RD_DOTA_UC} (a) and (b) depict the RD curves of the proposed LDM-RSIC and the comparison algorithms on the testing sets of DOTA and UC-M, respectively. The results show that the overall performance of LDM-RSIC outperforms all other methods, including traditional and learning-based algorithms, on both datasets, with the exception of CDC ($\rho=0.9$). We employ PSNR, MS-SSIM, and LPIPS metrics to measure the distortion of the decoded images. On both datasets, our RD curves outperform the comparison algorithm almost overall in terms of PSNR and MS-SSIM metrics, which indicates that LDM-RSIC achieves higher compression efficiency. 

Furthermore, while CDC ($\rho \! =\! 0.9$) shows competitive performance in perceptual quality (as reflected by low LPIPS values), it struggles to achieve competitive PSNR values on both datasets. This suggests that CDC ($\rho\!=\!0.9$) introduces synthetic, fake textures into the decoded images. In contrast, the proposed LDM-RSIC maintains high PSNR and achieves impressive LPIPS scores, demonstrating its balanced performance in both compression and perceptual quality.

Additionally, it is worth highlighting that JPEG2000*, when combined with the proposed scheme, delivers performance comparable to BPG, further showcasing the effectiveness of our method.

\subsection{Qualitative Comparison}
\begin{figure*}[!htbp]
	\centering
	\includegraphics[width=0.98\textwidth]{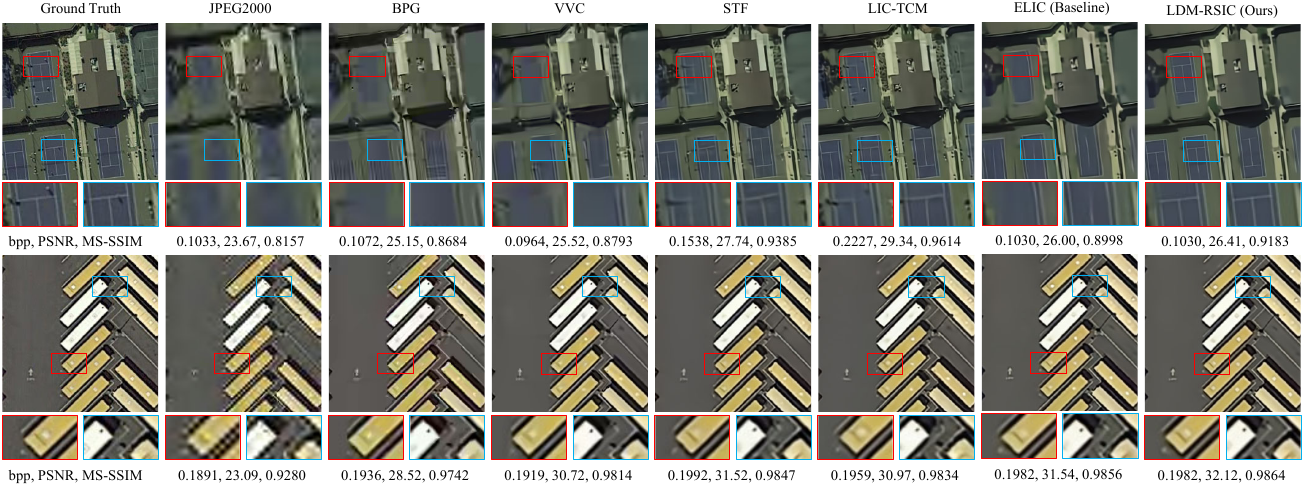}
	\caption{Compressed images by several compression algorithms on the testing images ``P0253" and ``P0006" of the DOTA testing set.}
	\label{Fig:vis_DOTA}
\end{figure*}

\begin{figure*}[!htbp]
	\centering
	\includegraphics[width=0.98\textwidth]{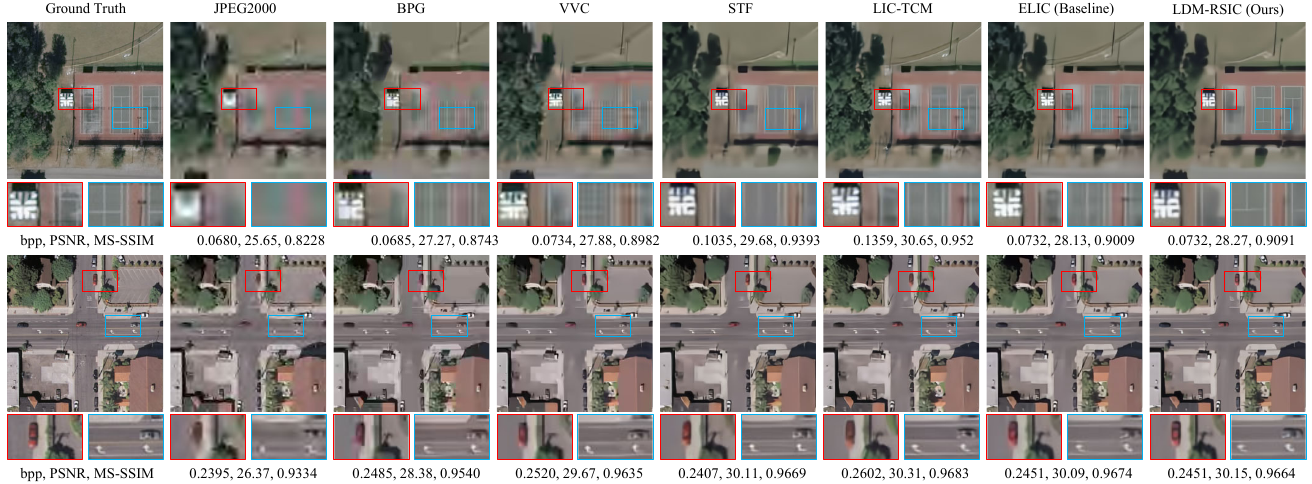}
	\caption{Compressed images by several compression algorithms on the images ``tenniscourt91" and ``intersection97" of the UC-M testing set.}
	\label{Fig:vis_UC}
\end{figure*}

To visualize the performance of these image compression algorithms, Figs. \ref{Fig:vis_DOTA} and \ref{Fig:vis_UC} depict the decoded images ``P0253", ``P0006", ``tenniscourt91", and ``intersection97" from the testing sets of DOTA and UC-M. At low bitrates, as illustrated in the image ``P0253" of Fig. \ref{Fig:vis_DOTA}, artifacts and artifacts from traditional algorithms like JPEG2000 and BPG significantly affect the quality of the decoded images, particularly at low bitrates. It is worth noting the comparison with the state-of-the-art conventional compression algorithm, VVC. Although the bpp of VVC is slightly lower than that of LDM-RSIC, there is a significant difference in the PSNR values of the two, with VVC being 0.89 dB lower than LDM-RSIC. In addition, it is clear that VVC is not able to adequately reconstruct the gridlines, whereas LDM-RSIC can clearly retain the edge information as well as have better contouring, demonstrating its superior performance. \par

When compared with learning-based algorithms, LDM-RSIC is observed to recover more details even at lower bitrates. For example, although the bpp of STF and LIC-TCM is 1.49 and 2.16 times higher than that of LDM-RSIC, respectively, LDM-RSIC demonstrates better recovery of the global structure of the images. Furthermore, at high bitrates, as shown in the image ``P0006" of Fig. \ref{Fig:vis_DOTA},  one can see that LDM-RSIC obtains 1.40 dB PSNR gains than VVC with similar values of bpp, and the edge details and structure information of LDM-RSIC are more impressive than VVC. Compared to the competitive learning-based image compression algorithm LIC-TCM, it can be observed that LDM-RSIC derives a 1.15 dB  PSNR increase, and the edge information has been better preserved with comparable bitrate.  \par

Moreover, as depicted in Fig. \ref{Fig:vis_UC}, despite VVC having a slightly higher bpp than LDM-RSIC, it fails to decode the structure of the windows. Notably, even though LIC-TCM has a bpp 1.86 times higher than LDM-RSIC, it still struggles to adequately recover the window structure. In the image ``intersection97", other methods struggle to clearly recover the red vehicles even at higher bpp. Although LIC-TCM exhibits higher PSNR and MS-SSIM scores than LDM-RSIC with higher bitrate, the road and vehicles appear more blurred. Therefore, we can safely demonstrate that the proposed LDM-RSCI presents a superior ability to decode texture-rich images compared to traditional and learning-based competitive image compression algorithms.

\begin{figure}[!htbp]
	\centering
	\includegraphics[width=0.48\textwidth]{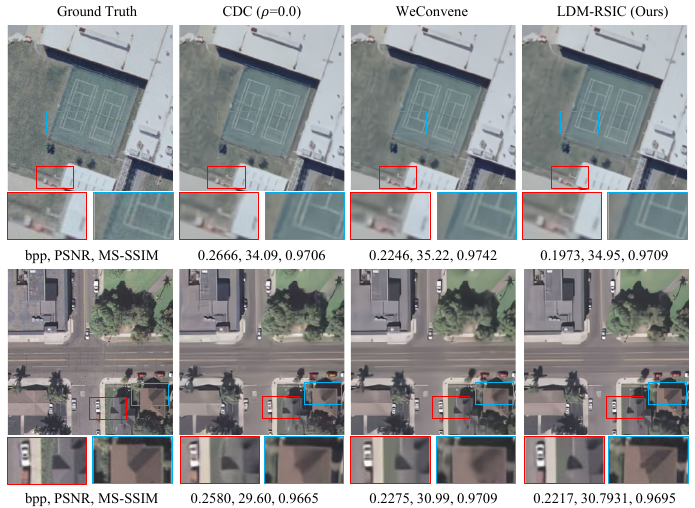}
	\caption{Visualization of the decoded images ``tenniscourt84” and ``intersection94” using state-of-the-art compression methods and the proposed LDM-RSIC.}
	\label{Fig:VIS_compare_UC}
\end{figure}

Additionally, Fig. \ref{Fig:VIS_compare_UC} provides visual comparisons of the decoded images from CDC ($\rho\!=\!0.0$), WeConvene, and the proposed LDM-RSIC. The results show that, despite consuming fewer bpp and achieving lower PSNR values, LDM-RSIC recovers significantly more edge details. Notably, although CDC ($\rho\!=\!0.0$) uses a higher bpp, both its PSNR and perceptual quality are inferior to those of LDM-RSIC. This strongly underscores the superior capability of the proposed method in enhancing perceptual quality.

\subsection{Prior Information Analysis}
Here, we focus on an analysis of the prior information of the developed LDM-RSIC and JPEG2000* from the two stages, namely the learned prior information in Stage I and the generated prior information in Stage II, to provide insights for future research directions. As outlined in Section \ref{Sec:Method}, Stage I is designed to learn the compression distortion prior conditioned on both the ground truth and the decoded image, while Stage II focuses on leveraging LDM to generate the prior solely conditioned on the decoded image. The closer the generated prior in Stage II is to the prior learned in Stage I, the more enhanced results we can obtain. \par
\begin{figure}[!htbp]
	\centering
	\subfigure[\fontsize{7}{10}\selectfont Prior analysis of LDM-RSIC]{
		\centering
		\begin{minipage}[t]{0.23\textwidth}
			\includegraphics[scale=0.58]{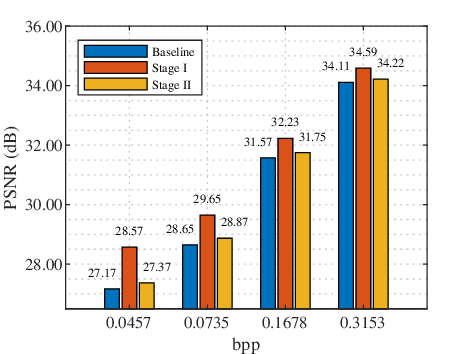}
		\end{minipage}	
	}%
	\subfigure[\fontsize{7}{10}\selectfont Prior analysis of JPEG2000*]{
		\centering
		\begin{minipage}[t]{0.23\textwidth}
			\includegraphics[scale=0.58]{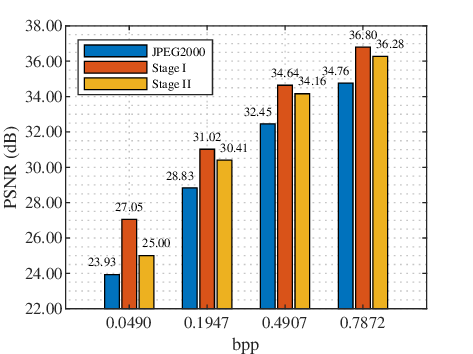}
		\end{minipage}	
	}%
	\centering
	\caption{PSNR performance of the proposed LDM-RSIC and JPEG2000* on Stages I and II with different values of bpp on the DOTA testing set.}
	\label{Fig: stage_prior}
\end{figure}
We refer to ELIC, the learning-based compression algorithm utilized in the proposed LDM-RSIC, as the baseline. Figs. \ref{Fig: stage_prior} (a) and (b) illustrate the PSNR of the proposed LDM-RSIC and JPEG2000* in Stages I and II with varying values of bpp. 
Take Fig. \ref{Fig: stage_prior}(a) as an example, it is evident that, under each bpp, the PSNR values of Stage I exhibit the highest results. Although the PSNR values of Stage II are higher than the baseline, the gap between Stages I and II cannot be considered negligible. For instance, under $ \text{bpp}=0.0457 $, the PSNR values of the baseline, Stage I, and Stage II are 27.17 dB, 28.57 dB, and 27.37 dB, respectively. This indicates that the gap between the two stages is as high as 1.20 dB, suggesting that LDM still has great potential to improve the generated prior information and narrow this gap. Fig. \ref{Fig: stage_prior}(b) shows the same result. Hence, exploring more competitive LDM techniques is expected to further enhance the RD performance of the proposed LDM-RSIC and JPEG2000*.

\subsection{Ablation Studies}
In this section, we investigate the effects of different designs of the proposed method. We conduct all experiments on the DOTA testing set. Here, we first explore the value of the diffusion step $ T $, then verify the validity of the LDM-based generated prior, and finally confirm the effectiveness of the developed LDM-based scheme in improving the compression efficiency of learning-based and traditional image compression algorithms.

\subsubsection{Diffusion Step}
\begin{figure}[!htbp]
	\centering
	\includegraphics[width=0.32\textwidth]{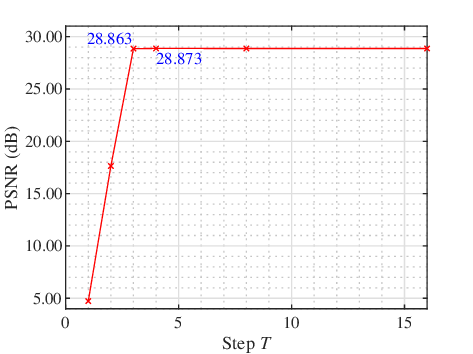}
	\caption{Ablation study of the total diffusion step $T$ of the proposed LDM-RSIC on the testing set of DOTA.}
	\label{Fig: Step_T}
\end{figure}

To explore the effect of the diffusion step $T$ in the LDM on the performance of the proposed LDM-RSIC, we vary the total number of iteration steps of the LDM-RSIC and adjust the parameter $\eta_t$ in Eq. (\ref{Eq:Add_t_noise})  to ensure that the $\mathbf{F}_T$ becomes Gaussian noise after the diffusion process, where $\mathbf{F}_T \sim \mathcal{N}(0, \mathbf{I})$.
Fig. \ref{Fig: Step_T} illustrates the average PSNR of the proposed LDM-RSIC at different values of $T$. The results indicate a significant improvement in the performance of LDM-RSIC as the step is increased to 3. Once $T$ reaches a value of 4, the performance of LDM-RSIC stabilizes.  Considering the trade-off between time complexity and performance, we set the total diffusion step $T$ to 4 in this paper. Moreover, our LDM-RSIC enjoys faster convergence speed compared to traditional DMs, which typically require more than 100 iterations. This accelerated convergence can be attributed to the deployment of the DM exclusively on low-dimensional latent spaces. \par

\subsubsection{Effectiveness of Prior Information}
To verify whether the prior information introduced by DFAM benefits the enhancement processing of LDM-RSIC, we conduct experiments on images ``P0216" and ``P0253" from the DOTA testing set under different values of $\lambda$. Table \ref{tab:aba_LRM} presents the results of the proposed LDM-RSIC with and without DFAM. As shown in the results, with DFAM, the PSNR gains are approximately 0.2 dB higher than the baseline in all cases of $\lambda$, indicating that LDM indeed generates useful prior information to improve decoding performance. \par
\begin{table}[htbp]
	\centering
		\renewcommand\arraystretch{1.1}
	\caption{Decoding performance of the proposed LDM-RSIC with and without DFAM on the DOTA testing set}
	\begin{tabular}{ccccc}
		\toprule
		$\lambda$ & Metrics & Baseline & w/o DFAM & w/ DFAM \\
		\hline
		\multirow{3}{*}{$4 \times 10^{-4}$} & PSNR $\uparrow$ & 27.17 & 27.25 & 27.37 \\
		& MS-SSIM $\uparrow$ & 0.8646 & 0.8663 & 0.8685 \\
		& LPIPS $\downarrow$ & 0.4135 & 0.4111 & 0.405 \\
		\multirow{3}{*}{$8 \times 10^{-4}$} & PSNR $\uparrow$ & 28.65 & 28.78 & 28.87 \\
		& MS-SSIM $\uparrow$ & 0.9006 & 0.9024 & 0.9036 \\
		& LPIPS $\downarrow$ & 0.3661 & 0.3657 & 0.3594 \\
		\bottomrule
	\end{tabular}%
	\label{tab:aba_LRM}%
\end{table}%

\begin{figure}[!htbp]
	\centering
	\includegraphics[width=0.45\textwidth]{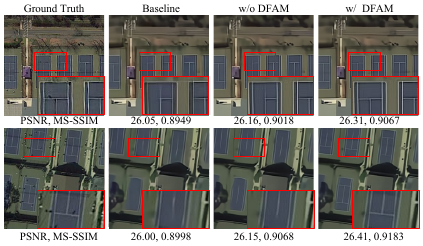}
	\caption{Visualization of the decoded images ``P0216" and ``P0253" of the baseline and the proposed LDM-RSIC with (w/) and without (w/o) DFAM. }
	\label{Fig:Abla_DFAM}
\end{figure}

Furthermore, to investigate whether the LDM-based generated prior contributes to recovering additional image details, Fig. \ref{Fig:Abla_DFAM} presents the decoded images with different model components under $\lambda=8\text{e}-4$. From the results, it is evident that without DFAM, the texture in the decoded image ``P0216" appears more blurred as well as crippled. Although there is a 0.11 dB PSNR gain compared to the baseline, it fails to generate additional edge details. In contrast, with DFAM, the decoded images exhibit more impressive edge texture and better structural information. The image ``P0253" further demonstrates these results. \par

In summary, we can demonstrate the effectiveness of the generated prior in providing more detailed information, thus enhancing the RD performance of RS image compression algorithms.

\subsubsection{Effectiveness of LDM-based Scheme}
In order to evaluate the effectiveness of the LDM-based scheme, we show the visualization of the decoded images obtained from LDM-RSIC and JPEG2000* obtained by employing the proposed LDM-based scheme to enhance the learning-based image compression algorithm ELIC and the traditional compression algorithm JPEG2000, respectively.

\subsubsection*{Enhancing ELIC} 
\begin{figure}[!htbp]
	\centering
	\includegraphics[width=0.45\textwidth]{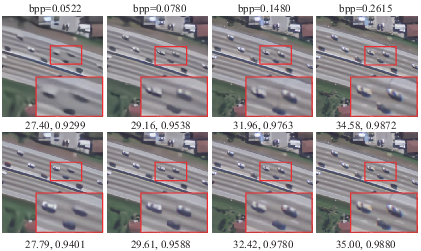}
	\caption{Visualization of the decoded images of the baseline (1st row) and the proposed LDM-RSIC (2nd row). PSNR and MS-SSIM are used for quality evaluation.}
	\label{Fig:com_base_our}
\end{figure}

We compare the baseline with the proposed LDM-RSCI, and visual results are depicted in Fig. \ref{Fig:com_base_our}. Notably, LDM-RSCI yields finer texture details compared to the baseline across various bpp values. For instance, at low bitrates (\textit{e.g.}, $ \text{bpp}=0.0522 $), the baseline's decoded vehicles appear blurry, whereas LDM-RSIC successfully restores the structure information. At higher bitrates (\textit{e.g.}, $ \text{bpp}=0.2615 $), LDM-RSIC not only enhances vehicle contour clarity but also recovers road markings.\par
\subsubsection*{Enhancing JPEG2000} 
\begin{figure}[!htbp]
	\centering
	\includegraphics[width=0.45\textwidth]{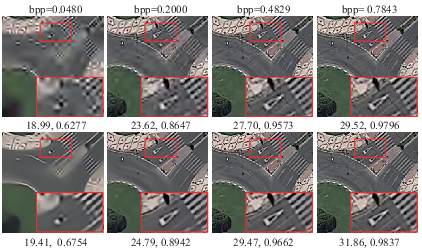}
	\caption{Visualization of the decoded images of the JPEG2000 (1st row) and the proposed JPEG2000* (2nd row).}
	\label{Fig:com_jpeg2000_enhance}
\end{figure}

As depicted in Fig. \ref{Fig:com_jpeg2000_enhance}, JPEG2000 decoded images exhibit noticeable artifacts. However, employing the LDM-based scheme significantly enhances texture detail clarity. For instance, at low bitrates (\textit{e.g.}, $ \text{bpp}=0.2000 $), JPEG2000 decoded image exhibits severe artifacts, whereas the enhanced version JPEG2000* enhances the image, resulting in a 1.17 dB increase in PSNR. This enhancement notably clarifies road markings. Similarly, at higher bitrates, the structure of the images recovered with JPEG2000* is significantly clearer. \par
In conclusion, the experiments demonstrate the positive impact of the LDM-based scheme on both deep learning-based and traditional image compression algorithms for decoding high-quality images.

\subsection{Model Complexity}
The complexity of the proposed LDM-RSIC model is evaluated in comparison to other state-of-the-art image compression algorithms, as shown in Table \ref{tab:parameter_FLOP}. The DOTA testing set is utilized for the experiment. The complexity is measured in terms of FLOPs (floating-point operations), the number of parameters, and encoding/decoding times on both CPU and GPU platforms.  \par
\begin{table*}[!htbp]
	\renewcommand\arraystretch{1.1}
	\centering
	\caption{Complexity Comparison of several learning-based image compression algorithms}
	\begin{tabular}{ccccccc}
		\toprule
		\multirow{2}{*}{Methods} &  \multirow{2}{*}{FLOPs (G)} & \multirow{2}{*}{Parameters (M)} & \multicolumn{2}{c}{CPU} & \multicolumn{2}{c}{GPU} \\
		\cline{4-7}        &     &     & Encoding Time (s) & Decoding Time (s) & Encoding Time (s) & Decoding Time (s) \\
		\hline
		Entroformer \cite{qian2022entroformer} &  44.76 & 12.67 & 1.3482  & 0.5018  & 0.2762  & 0.0919  \\
		STF \cite{zou2022devil} & 99.83 & 33.35 &  1.1395  & 1.3175  & 0.1343  & 0.1879  \\
		ELIC \cite{he2022elic} & 31.66  & 54.46 & 1.7934 & 1.6796 & 0.4671 & 0.2507 \\
		LIC-TCM \cite{liu2023learned} &  35.23 & 44.97 & 22.0959  & 21.0951  & 0.4637  & 0.4045  \\
		WeConvene \cite{fu2024weconvene} & 150.76 &  105.51 & 50.6482 & 49.3341 & 0.3472 & 0.3880 \\	
		LDM-RSIC (Ours) &  94.41 & 79.53 & 1.7934  & 1.8530  & 0.4671  & 0.4413 \\		
		\bottomrule
	\end{tabular}%
	\label{tab:parameter_FLOP}%
\end{table*}%

In terms of FLOPs, LDM-RSIC requires 94.41G, placing it in the higher range compared to lighter models like Entroformer (44.76G) but lower than the computationally expensive WeConvene (150.76G). Similarly, the number of parameters for LDM-RSIC is 79.53 M, which is significantly higher than Entroformer (12.67M) but still lower than WeConvene (105.51M). This indicates that while LDM-RSIC has a relatively complex architecture, it is not the most computationally heavy among the compared models.

For encoding and decoding times, LDM-RSIC performs efficiently on the GPU, with encoding and decoding times of 0.4671s and 0.4413s, respectively. These results show that while LDM-RSIC has higher computational requirements than lightweight models such as STF and Entroformer, it is still capable of producing high-quality compressed images with reasonable computational costs. On the CPU, the encoding and decoding times are 1.7934s and 1.8530s, respectively, which are comparable to other models with similar complexities.

In conclusion, LDM-RSIC demonstrates moderate-to-high computational complexity, reflecting its ability to capture more intricate image features compared to simpler models, while maintaining reasonable inference times, especially on GPU platforms. This trade-off between complexity and performance is appropriate for applications requiring high-quality image compression without excessive computational overhead.

\section{Conclusion} \label{Sec:Conclusion}
In this paper, we develop the LDM-RSIC to enhance the RD performance of the learning-based image compression algorithm ELIC. Specifically, LDM-RSIC utilizes the LDM to generate compression distortion prior, which is then integrated into the Transformer-based MEN to enhance the quality of the decoded images. Additionally, we propose a channel attention and gate-based DFAM to better utilize the prior. Furthermore, we apply the proposed LDM-based scheme to enhance the traditional image compression algorithm JPEG2000, significantly improving the perceptual quality of the decoded images.
Extensive experiments on two widely used RS image datasets demonstrate that LDM-RSIC significantly outperforms state-of-the-art traditional and learning-based image compression algorithms in terms of objective performance and subjective perception quality. 


\bibliographystyle{IEEEtran}
\bibliography{reference.bib}

\end{document}